\documentclass[showpacs,aps,prl,twocolumn,superscriptaddress,10pt]{revtex4-2}
\usepackage{graphicx} 
\usepackage{bm}
\usepackage{color}

\usepackage{amsmath}
\usepackage{amssymb}
\usepackage{enumerate}
\usepackage{xspace}
\usepackage{mathrsfs}
\usepackage{mathptmx}
\usepackage[colorlinks=true,linkcolor=blue,citecolor=blue,urlcolor=blue]{hyperref}
\usepackage{orcidlink}
\usepackage[utf8]{inputenc}

\begin{document}

\title{Negative temperature coefficient of Gilbert damping in magnetic bilayers}  

\author{Lulu Cao\orcidlink{0009-0005-7572-5780}}
\affiliation{Key Laboratory of Quantum Materials and Devices of Ministry of Education, School of Physics, Southeast University, Nanjing, 211189, China}
\affiliation{School of Physics, Engineering and Technology, University of York, York, YO10 5DD, UK}
\affiliation{National Key Laboratory of Spintronics, Nanjing University, Suzhou 215163, China}
\affiliation{School of Integrated Circuits, Nanjing University, Suzhou 215163, China}
\author{Yuting Gong\orcidlink{0000-0001-8087-4346}}
\affiliation{Jiangsu Provincial Key Laboratory of Advanced Photonic and Electronic Materials, School of Electronic Science and Engineering, Nanjing University, Nanjing 210093, China}
\author{Xianyang Lu\orcidlink{0000-0003-3815-4032}}
\affiliation{National Key Laboratory of Spintronics, Nanjing University, Suzhou 215163, China}
\affiliation{School of Integrated Circuits, Nanjing University, Suzhou 215163, China}
\author{Yongbing Xu\orcidlink{0000-0002-7823-0725}}
\affiliation{National Key Laboratory of Spintronics, Nanjing University, Suzhou 215163, China}
\affiliation{School of Integrated Circuits, Nanjing University, Suzhou 215163, China}
\affiliation{Jiangsu Provincial Key Laboratory of Advanced Photonic and Electronic Materials, School of Electronic Science and Engineering, Nanjing University, Nanjing 210093, China}
\author{Ya Zhai\orcidlink{0000-0002-9006-9575}}
\email{yazhai@seu.edu.cn}
\affiliation{Key Laboratory of Quantum Materials and Devices of Ministry of Education, School of Physics, Southeast University, Nanjing, 211189, China}
\author{Jing Wu\orcidlink{0000-0003-0007-4328}}
\email{jing.wu@york.ac.uk}
\affiliation{School of Physics, Engineering and Technology, University of York, York, YO10 5DD, UK}
\author{Roy~W.~Chantrell\orcidlink{0000-0001-5410-5615}}
\affiliation{School of Physics, Engineering and Technology, University of York, York, YO10 5DD, UK}
\author{Richard~F.~L.~Evans\orcidlink{0000-0002-2378-8203}}
\email{richard.evans@york.ac.uk}
\affiliation{School of Physics, Engineering and Technology, University of York, York, YO10 5DD, UK}
\begin{abstract}
The Gilbert damping of magnetic materials is an important magnetic parameter that determines the switching speed and energy dissipation of spintronic devices. In simple metals, the intrinsic Gilbert damping increases with temperature and diverges near the Curie temperature as a result of spin fluctuations. Here we present atomistic simulations and experimental measurements showing surprising and opposite behavior in Py/Nd bilayers, where the Gilbert damping \emph{decreases} with increasing temperature. The effect arises because of the enhanced damping at the interface as a result of spin pumping, where elevated temperatures cause a dynamic separation of the interfacial and bulk magnetization during relaxation. Furthermore, the temperature dependence of the damping can be controlled by varying the thickness of the Nd capping layer. Our findings present a new spintronic effect that can be used to modify the dynamic properties of nanoscale materials and devices for enhanced energy efficiency or with improved switching dynamics.
\end{abstract}

\maketitle
The Gilbert damping factor is an important magnetic parameter that determines the speed of magnetization dynamics in spintronics devices, such as magnetic sensors and magnetic random access memory. The intrinsic Gilbert damping arises at the electronic structure level due to magnon/electron~\cite{Hickey2009,Schoen2016,Lu2024}, or magnon/phonon interactions in insulators. However, in nanoscale devices extrinsic contributions including magnon/magnon interactions and scattering at surfaces become important. Theoretically, several models have attempted to explain the origin of damping, such as the $s-d$ exchange scattering model, the Fermi surface breathing model, the two-magnon scattering model, and the spin pumping model\cite{Zhao_2016,PhysRevB.66.224403,PhysRevLett.88.117601,MankovskyPRB2019}. For interfacial Gilbert damping that arises between ferromagnetic / nonmagnetic bilayers, the spin pumping effect is widely considered to be responsible for the dependence of the damping on the thickness of the non-magnetic layer~\cite{Liu2014}. However, most studies focused on the variation of Gilbert damping values at room temperature or lower, with few research groups investigating changes in Gilbert damping over a wider temperature range including close to the Curie temperature. 
In thin films of permalloy the temperature dependence of the damping was found to be non-monotonic for small film thicknesses, with a peak around $T = 50$~K~\cite{Zhao_2016} but weakly temperature dependent with a positive temperature coefficient for thicker films. The non-monotonic behavior was attributed to a potential spin-reorientation transition at low temperature due to enhanced interfacial anisotropy~\cite{Zhao_2016}. The microstructure is also known to have an important effect on Gilbert damping and also its temperature dependence with recent studies on FePt~\cite{Richardson2018} finding a decrease in the ferromagnetic resonance (FMR) linewidth at higher temperatures, which is often interpreted as a decrease in Gilbert damping. However, recent theoretical calculations attributed the effect to inhomogeneous line broadening at low temperatures (increasing the effective damping) that reduces at high temperatures~\cite{Strungaru2020}.

As well as its fundamental importance, the Gilbert damping parameter is especially important for devices including magnetic random access memory (MRAM)~\cite{Whitney2023,Thomas2018}, heat assisted magnetic recording (HAMR) and magnetic hyperthermia. In MRAM devices low damping can aid switching speeds and is a desirable property for fast switching. However, Joule heating can cause an increase in device temperature during switching that usually increases the effective damping constant. Similarly for HAMR devices, dynamic heating can cause changes in damping that are important for the switching probability and thermal stability of magnetic grains~\cite{Strungaru2020}. The Gilbert damping and its temperature dependence is therefore an important engineering parameter for devices. 

 
In this letter, we report the temperature dependence of the Gilbert damping of a Py/Nd bilayer using an atomistic spin dynamics model including spin pumping and  complementary time-resolved magneto-optical Kerr effect (TR-MOKE) measurements of the pump fluence dependence of Gilbert damping. 
To study the temperature effect on Gilbert damping we first consider atomistic spin dynamics simulations of the magnetic relaxation process at a range of temperatures $T = 20$-$800$ K. Based on a spin accumulation model\cite{Cao_2024}, we introduce bulk and interface magnetic damping to investigate the thickness dependence of the damping in Py(6 nm)/Nd($t$ nm) bilayer films \cite{Barati2015,Zhao_2016,Liu2014,Cao_2024}, as shown in Figure 1. The simulations utilize a classical spin model with a 3D Heisenberg spin Hamiltonian\cite{Evans2014} describing the energy contribution of the magnetic systems given by:
\begin{equation}
\mathcal{H}=- \sum\limits_{i < j}{{J_{ij}}{{{\bf{S}}}_i}\cdot
{{{\mathbf{S}}}_j}}-{k_{\mathrm{u}}} \sum\limits_{i}({{{\mathbf{S}}_i}}\cdot{{\bf{e}}})^2- {\lvert\mu_{\mathrm{s}}\rvert}\sum\limits_{i}{\mathbf{S}}_i \cdot{{\bf{B}}_{\mathrm{app}}},
\label{eq:Ham}
\end{equation}
 where $J_{ij}$ is the nearest neighbor exchange integral between spin sites $i$ and $j$. Nearest neighbor exchange is used following Hinzke et al.\cite{Hinzke2015} who showed using first principle calculations that direct exchange was dominant. $\mathbf{S}_{i}$ is the local normalized spin moment, $\mathbf{S}_{j}$ is the normalized spin moment of the neighboring atom at site $j$, $k_{\mathrm{u}}$ is the uniaxial anisotropy constant, $\bf{e}$ is the unit vector of the easy axis. Each magnetic moment is normalized so that $\mathbf{S}=\frac{\boldsymbol{\mu}}{\lvert \mu_{\mathrm{s}}\rvert}$, where $\lvert \mu_{\mathrm{s}}\rvert$ is the magnitude of the spin moment. ${\bf{B}}_{\mathrm{app}}$ is the applied external field. The magnetic parameters of Py and Nd are listed in the Supplementary Information (SI) Tab~I ~\cite{supp} with parameters derived from previous publications\cite{Sampanapai2019,Callen1966TheLaw,PhysRev.96.1335,10.1063/1.1713369,Irkhin1988,10.1063/1.1713369,Irkhin1988} 
\begin{figure}
    \centering
    \includegraphics[width=1.0\linewidth]{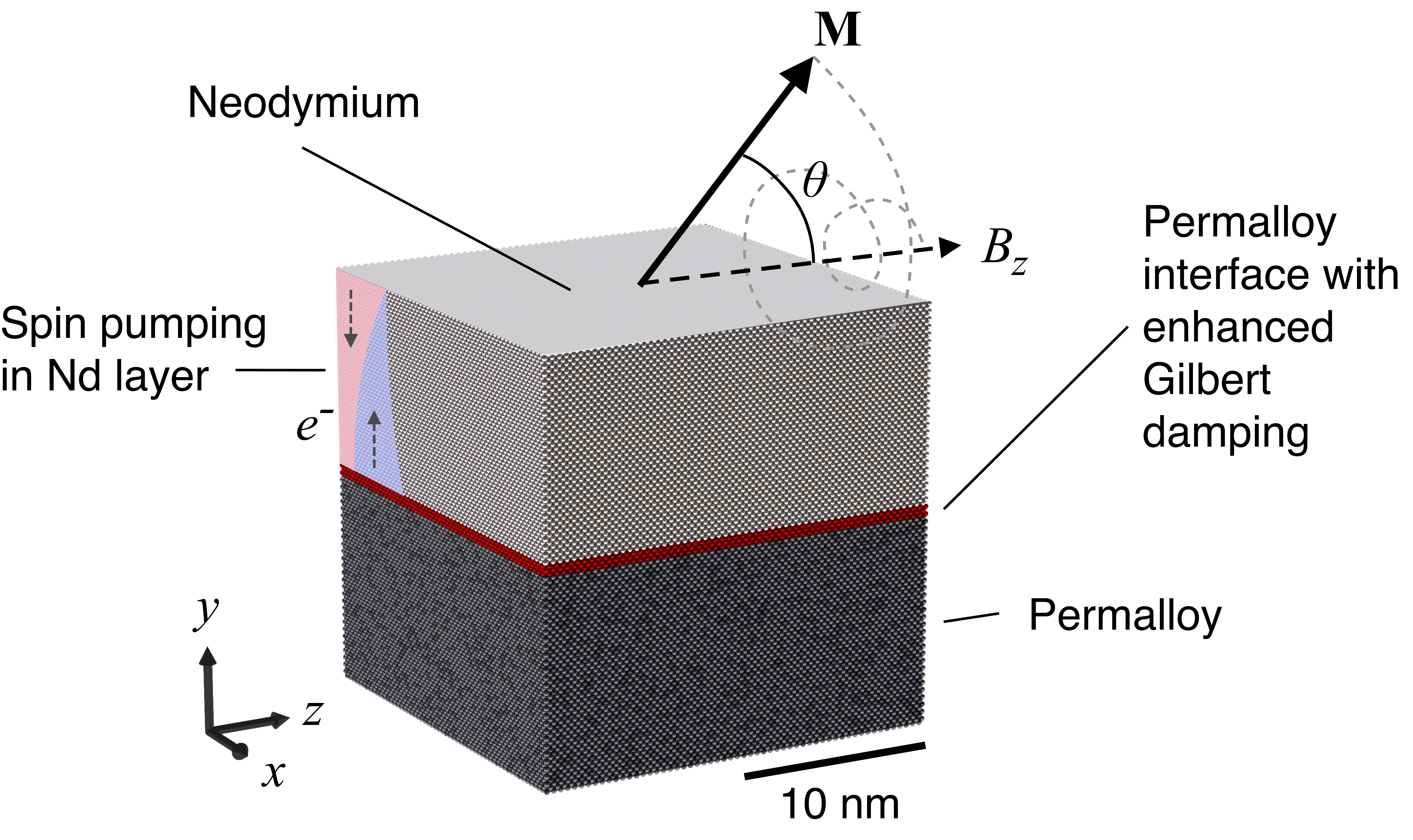}
    \caption{Schematic view of the simulated Py (6 nm)/Nd (t nm) bilayer. The structural dimensions are $\textit{20}\times\textit{20}\times\left(\textit{6}+\textit{t}\right) \textit{nm}^{3}$, where the dark  grey underlayer and the red inserted layer represent the bulk and interface Py layer, respectively. The light-grey top layer represents the Nd layer.}
    \label{fig:structures}
\end{figure}
We simulate magnetic dynamic behavior using the stochastic Landau–Lifshitz-Gilbert (sLLG) equation applied at the atomistic level~\cite{vampire,EllisLTP2015} and given by
\begin{equation}
\frac{\partial {\mathbf{S}_{i}}}{\partial t} = -\frac{\gamma }{1 + \alpha^2}\left[{\mathbf{S}_{i}}\times \mathbf{B}_{\mathrm{eff}}^i + {\alpha} {\mathbf{S}_{i}}\times ({\mathbf{S}_i}\times  \mathbf{B}_{\mathrm{eff}}^i )\right],
\end{equation}
where $\mathbf{S}$ is the normalized spin moment, $\gamma$ is the absolute value of the gyromagnetic ratio, {$\alpha$} is the intrinsic damping constant which represents the coupling of the spin to the heat bath. {$\mathbf{B}_{\mathrm{eff}}^i = \mathbf{B}_\mathrm{th}^i-\frac{1}{|\mu_S|}\frac{\partial \mathscr{H}}{  \partial \mathbf{S}_{i}}$} is the effective field that acts on the local spin moment determined from the Heisenberg spin Hamiltonian. The effective field is enhanced by a thermal fluctuating field $\mathbf{B}_{\text{th}}^i\left(t\right) =\Gamma\left(t\right)\sqrt{\frac{2{\alpha} k_{B}T}{\gamma\mu_{s}\mathrm{\Delta}t}}$ within a Langevin Dynamics formalism~\cite{Berkov1993} where \(\Gamma\left(t\right)\) is the Gaussian distribution and $T$ is the temperature. The intrinsic Gilbert damping ${\alpha}$ is treated as a zero-temperature constant for the simulations, and is different for the bulk and interface layer. For increasing Nd layer thickness we model the effect of the spin pumping by increasing the intrinsic damping at the Nd/Py interface ${\alpha}_\mathrm{int} = {\alpha_{\max}}\left( 1 - e^{\frac{- 2d_{\text{NM}}}{\lambda_{\text{sdl}}}} \right)$, with $d_{\text{NM}}$ the thickness of the Nd layer and $\lambda_{\text{sdl}}$ the spin diffusion length\cite{10.1063/1.1853131,Cao_2024}. More details of the spin pumping implementation and spin-pumping induced interface damping values in the simulation are presented in supplementary information~\cite{supp}.

\begin{figure}[!tb]
    \centering
    \includegraphics[width=1.0\linewidth]{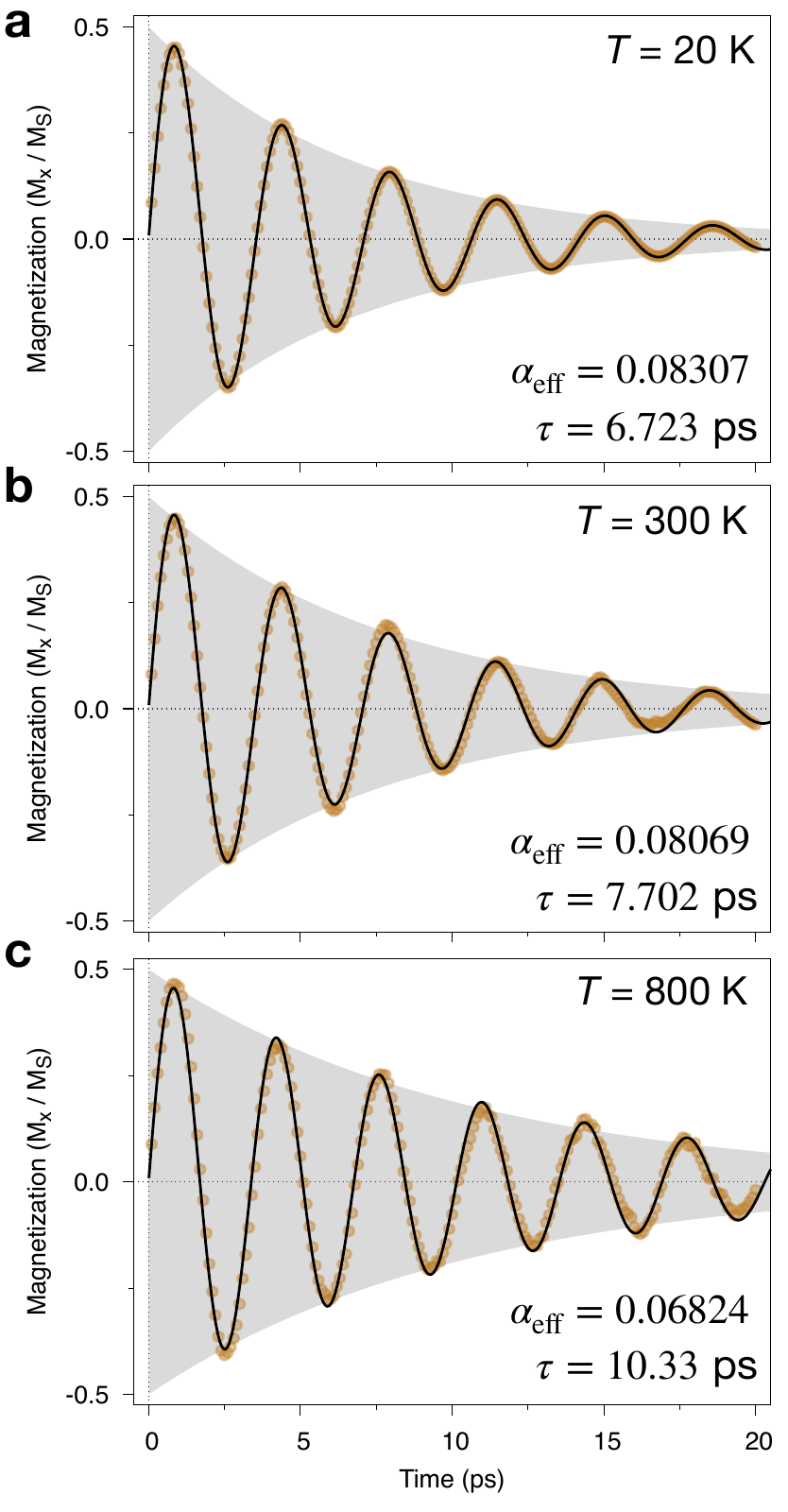}
    \caption{Simulated time evolution of the magnetisation trace (points) for the Py (6 nm)/Nd (32 nm) bilayer at different temperatures (a) $T = 0$ K, (b) $T = 300$ K, (c) $T = 800$ K showing a systematic \textit{decrease} in the effective Gilbert damping $\alpha_\mathrm{eff}$ with increasing temperature as seen from the amplitude of the envelope function depicted by the shaded region. {The lines are fitted to the data using Equation~\ref{eq:relax1}.}}
    \label{fig:relaxation}
\end{figure}

\begin{figure}[!tb]
    \centering
    \includegraphics[width=1.0\linewidth]{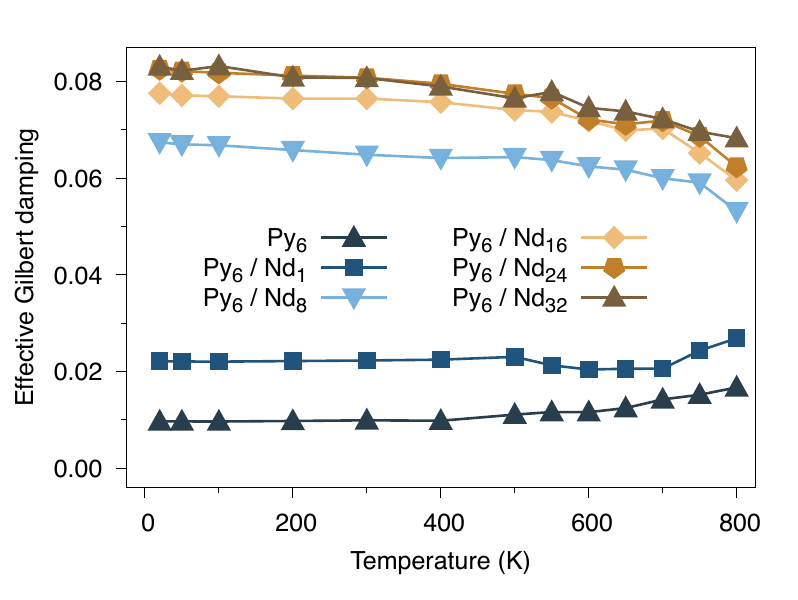}
    \caption{Simulated mean-effective Gilbert damping as a function of temperature for Py (6 nm)/Nd (0, 1, 8, 16, 24, 32 nm) bilayers. The data show a consistent decrease of the effective damping with temperature for Py (6 nm)/Nd (8, 16, 24, 32 nm) bilayers, and the expected increase for the case for pure Py.}
    \label{fig:damping}
\end{figure}

To calculate the effective Gilbert damping $\alpha_\mathrm{eff}$, we applied a $B_z = 10$ T external magnetic field along the $z$-direction\cite{Sampanapai2019}. The magnetisation is initially oriented in the $yz$ plane at an angle of 30$^\circ$ to the applied field from where it undergoes damped precession into the field direction, as shown in Fig.~\ref{fig:structures}.  Different angles between the initial magnetisation direction and the applied field are simulated for Py (6 nm)/Nd(32 nm) and are shown in the Supplementary Information~\cite{supp}. Fig.~S1 shows similar results independent of the initial magnetisation angle. 
The time evolution of the $x$ component of magnetisation for the Py(6 nm)/Nd(32 nm) bilayer in the atomistic model at different temperatures is shown in Fig.~\ref{fig:relaxation}. For the Py/Nd bilayer, as the net magnetisation of the system evolves back to the equilibrium state, the precession frequency and the relaxation time are extracted using the equation,
\begin{equation}    
M\left( t \right) = A\exp\left( - \frac{t}{\tau} \right)\sin\left( 2\pi ft \right),
\label{eq:relax1}
\end{equation}
where $A$ is the amplitude of the magnetisation, $f$ is the precession frequency and  $\tau$ is the relaxation time. From Eq.~\ref{eq:relax1}, the relaxation time of the magnetisation at different temperatures is extracted and the effective Gilbert damping ($\alpha_\mathrm{eff}$) can be determined using equation {$\alpha_\mathrm{eff}=1 / 2\pi\text{f}\tau$}. To obtain statistically accurate results, we performed simulations with 28 different numerical random-seeds for each parameter set, and then averaged the values and calculated the variance to obtain the mean-effective Gilbert damping, as shown in the Supplementary Information Fig. S2 ~\cite{supp}. The mean-effective Gilbert damping for Py/Nd bilayers with different Nd thickness was calculated as a function of temperature and is shown in Fig.~\ref{fig:damping}. In agreement with previous work \cite{Zhao_2016} we can see that for a pure Py thin film the Gilbert damping value increases with temperature as expected from increased thermal fluctuations. In contrast, the Py/Nd(32 nm) bilayers show a remarkable \textit{decrease} in the effective damping with increasing temperature, with an effective Gilbert damping value of $\alpha_\mathrm{eff} = 0.0822$ at $T = 20$ K, which decreases to $\alpha_\mathrm{eff} = 0.0636$ at $T=800$ K. This is an unexpected result as higher temperatures induce larger thermal fluctuations in the magnetisation that leads to a slightly slower relaxation towards the equilibrium direction. Here, the damping has an additional surface contribution from the spin pumping, which clearly plays a role on the effective damping, especially at higher temperatures.

In order to verify the predictions of the atomistic simulations, we performed experimental measurements of the temperature dependence of the Gilbert damping for the magnetic bilayer using TR-MOKE with variable pump fluence to change the effective temperature during relaxation. The structure and magnetic properties of the bilayer films have been presented in previous work~\cite{Cao:23}. To study the effect of temperature on the effective Gilbert damping, ultrafast laser pulses are used to excite magnetisation dynamics and elevate the temperature of the system. The ultrafast demagnetisation behavior modifies the anisotropy field and magnetisation within a few picoseconds; then the anisotropy field recovers and the magnetisation returns to the initial equilibrium position via a precession process which will last for hundreds of picoseconds. The schematic structure of the sample and the geometry of the TR-MOKE measurement are shown in detail in Figure S3(a). To quantitatively analyze the data, we use a phenomenological equation to fit the Kerr rotation trace:
\begin{equation}
    \Delta\theta = A\exp\left(\frac{-t}{\tau}\right)\cos(2\pi ft+\phi)+B(t)
    \label{eq:relax}
\end{equation}
where $A$, $\tau$, $f$ and $\phi$ are the magnetisation precession amplitude, the relaxation time, the precession frequency and the initial phase, respectively. $B(t)$ is the background term associated with the demagnetisation recovery process. Fitting the transient Kerr traces under different pump fluences with equation~\ref{eq:relax}, we obtain the precession frequency $f$ and the relaxation time $\tau$. In the experiment, we observed only one mode of precession frequency, which excludes the contribution of multi-mode spin waves to the effective Gilbert damping. Additionally, it is known that the effective Gilbert damping is stable at a high magnetic field~\cite{Cao:23}. Therefore, we can approximately take {into} account the effective Gilbert damping under a high magnetic field from the intrinsic damping of the material and the enhancement damping from the spin pumping effect. The effective electron temperature $T_e$ immediately after the pulse can be deduced from the absorbed laser energy per unit volume $E_a$ using~\cite{PhysRevB.78.174422,PhysRevB.36.2920,PhysRevLett.76.4250}
\begin{equation}
    E_a=\gamma(T_e^2-T_0^2)/2
    \label{eq:EA}
\end{equation}
where $T_0$ is the initial electronic temperature (room temperature). This arises from the fact that the electronic specific heat is linearly dependent on temperature such that $C_e= \gamma T_e$, with $\gamma_{Fe} =0.7 mJ/cm^3K^2$, $\gamma_{Ni} =6 mJ/cm^3K^2$. Meanwhile, $E_a$ can also be estimated according to the optical parameters:
\begin{equation}
    E_a=\left[1-\exp\left(-\frac{d}{\lambda}\right)\right]\frac{I_\mathrm{F}(1-R)}{d}
    \label{eq:EA-optical}
\end{equation}
where $I_\mathrm{F}$ is the incident laser fluence, $R$ is the reflectivity of the material at the wavelength, $\lambda$ is the absorption length and $d$ is the film thickness. Combining Eqs.~\ref{eq:EA} and \ref{eq:EA-optical}, we obtained the electronic temperature of Ni in the range between 540 K and 850 K, and the electronic temperature of Fe between 1350 K and 2350 K, with a weighted average electron temperature of NiFe between 580 and 920 K. The effective Gilbert damping and the dependence of the electron temperature with pump fluence are shown in Figure S3 (b)-(d). Using this, we can therefore characterize the temperature dependence of the effective Gilbert damping.

\begin{figure}
    \centering
    \includegraphics[width=1.0\linewidth]{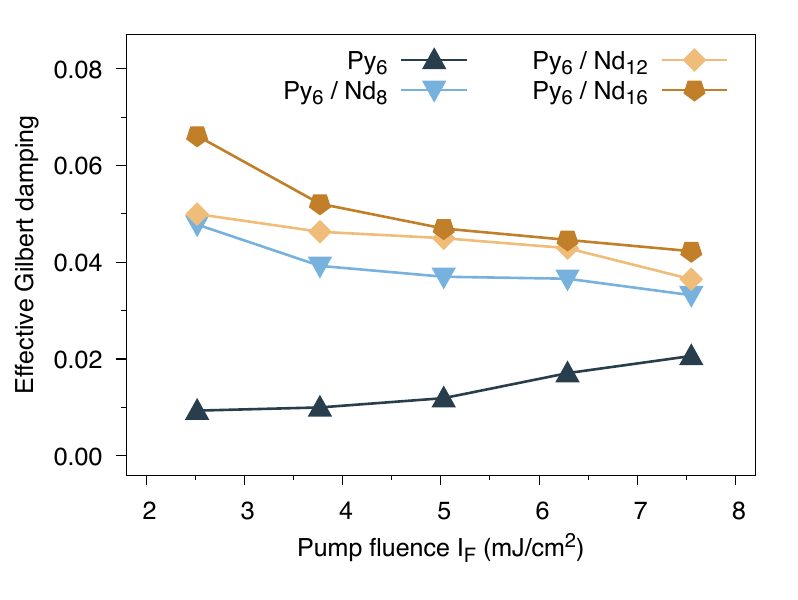}
    \caption{Experimentally measured effective Gilbert damping as a function of the pump fluence of the laser $I_\mathrm{F}$ for Nd ($t$ = 0, 8, 12, 16 nm)/Py (6 nm) bilayers.}
    \label{fig:structures_and_damping}
\end{figure}

The effective Gilbert damping values for Nd($t$ nm)/Py(6 nm) bilayer under different pump fluences, are shown in Fig~\ref{fig:structures_and_damping}. For a single-layer Py thin film, the effective Gilbert damping value continuously increases with increasing pump fluence due to the increasing presence of magnons with increasing temperature~\cite{10.1063/5.0056059,article}. However, for the Nd (8, 12, 16 nm)/Py bilayer, the effective Gilbert damping values decrease continuously with increasing pump fluence, as seen in the earlier atomistic simulations. Further details of the Py/Nd measurements are presented in Supplementary Fig.~S4 ~\cite{supp}.
%

To investigate the general effect of spin-pumping on temperature dependent damping, we also performed comparative measurements of the effective Gilbert damping values for Pt(2, 4 nm)/Py (6 nm) and {${Pt}_{0.68}{Nd}_{0.32}$/Py} bilayers under different pump fluence, shown in Fig.~S6. The TR-MOKE measurements show an increasing effective Gilbert damping with temperature as with the pure Py thin film. In Pt/Py ({${Pt}_{0.68}{Nd}_{0.32}$/Py}) bilayers the interface spin accumulation is very small\cite{Liu2014,doi:10.1126/sciadv.aat1670,10.1063/1.1853131} due to a short spin diffusion length of $0.609 \pm 0.092$ nm ($3.509 \pm 0.732$ nm), as measured in Fig.~S5. Thus the spin-pumping effect in Pt and ${Pt}_{0.68}{Nd}_{0.32}$ are already saturated for this thickness. This limits the enhancement of interfacial Gilbert damping due to spin pumping, meaning that the bulk and interface damping are close in value. This suggests that the origin of the effect is due to a small separation of the interfacial and bulk magnetisation dynamics, as shown in Fig.S7, where the surface experiences much larger thermal fluctuations due to the loss of exchange bonds through reduced coordination at the surface~\cite{NguyenPRB2022}. 


To evaluate this possibility, we simulate an idealized system where the spin-pumping enhanced damping is distributed much deeper into the Py film. In this computational experiment, we assume a bulk-layer and interface layer each 3 nm thick with different Gilbert damping constants. The magnetic parameters and the simulation results are shown in detail in Figure S8. The mean-effective Gilbert damping value is 0.01 at 300 K, which is consistent with previous results\cite{Cao_2024,Cao:23,Liu2014}. In this simulation, the mean-effective Gilbert damping value increases with increasing temperature, as for the pure Py film. Thus, the origin of the negative temperature-dependent coefficient of Gilbert damping can be ascribed to the strong localization of the damping enhancement to the interface Py region. Here the spin fluctuations are largest due to the reduction in coordination\cite{NguyenPRB2022}. This leads to a rapid loss of enhancement resulting in a decrease in the damping. This behavior originates from the large enhancement of the spin-pumping induced interface damping, combined with enhanced surface spin fluctuations leading to an increased surface thermal fluctuations of the magnetisation, as shown in Fig.S9 and Fig.S10~\cite{supp}. This partially decouples the surface and bulk magnetisation, thus reducing the effective damping enhancement due to the spin pumping. Finally, considering the case of small spin-pumping enhanced interface damping in the case of monolayer Nd and Pt layers, we see a small increase in effective damping as a function of temperature in Fig~\ref{fig:damping} for the case of 6Py/1Nd. In this case the bulk and interfacial damping are close in value, leading to a reduced surface effect.


In summary, the effective Gilbert damping dependence on temperature for Nd/Py bilayers is systematically studied using atomistic spin dynamics (ASD) simulations and time-resolved magneto-optical Kerr effect (TR-MOKE) measurements. When the interface spin accumulation is large as in the case of the Nd/Py bilayer the effective Gilbert damping shows that the coefficient of the temperature dependence is negative due to surface spin fluctuations separating the dynamics of the surface and bulk layers. Conversely, when interface spin accumulation is small as seen in Pt/Py and ${Pt}_{0.68}{Nd}_{0.32}$/Py bilayers the effective Gilbert damping shows a positive correlation with temperature due to the usual thermal fluctuations. The ability to engineer different temperature coefficients of Gilbert damping is particularly useful for deep-nanoscale spintronic devices, especially when operating at elevated temperatures. Other rare-earth metals may also exhibit similar effects, providing a rich optimization space for different materials and devices.

\section*{Author contributions}
Lulu Cao performed sample preparation, measurements, atomistic simulations, analyzed the data, and wrote the manuscript; Richard F. L. Evans and Roy W. Chantrell performed atomistic and spin accumulation calculations, analyzed the data, and modified the manuscript; Yuting Gong, Xianyang Lu, Yongbing Xu and Jing Wu performed the sample measurement. Ya Zhai, Jing Wu, Richard F. L. Evans and Roy W. Chantrell performed project administration and supervision.

\section{Data availability statement}
All data supporting the findings of this study are available online~\cite{data}.

\section{Conflict of interest}
The authors declare that they have no known competing financial interests or personal relationships that could have appeared to influence the work reported in this paper.

\section{acknowledgments}
This work is supported by the National Natural Science Foundation of China (Grant Nos. 52071079, 12274071, 12241403); the National Natural Science Foundation of China (Grant Nos. 12104216); the National Key R$\&$D Program of China (Grant Nos.2021YFB3601600). We gratefully acknowledge computational resources from the \textsc{viking} computer system used in this project, which is a high performance compute facility provided by the University of York.

\appendix

\bibliography{library}


\end{document}